\documentstyle[12pt,doublespace2,epsfig]{article}
\def\lapprox{\hbox{\lower .8ex\hbox{$\,\buildrel < \over\sim\,$}}}
\def\gapprox{\hbox{\lower .8ex\hbox{$\,\buildrel > \over\sim\,$}}}

\begin{document}

{\Large \bf No surviving evolved companions to the \\
progenitor of supernova SN~1006}

\bigskip\bigskip\noindent

\noindent
Jonay I. Gonz\'alez Hern\'andez$^{1,2}$, Pilar Ruiz-Lapuente$^{3,4}$,
Hugo M. Tabernero$^{5}$, David Montes$^{5}$, Ramon Canal$^{4}$,
Javier M\'endez$^{4,6}$, Luigi R. Bedin$^{7}$  

\bigskip\medskip

\noindent
$^{1}$ Instituto de Astrof\'{\i}sica de Canarias,
E-38205 La Laguna, Tenerife, Spain

\noindent
$^{2}$ Departamento de Astrof{\'\i}sica, Universidad de La Laguna, 
E-38206 La Laguna, Tenerife, Spain 

\noindent
$^{3}$ Instituto de F{\'\i}sica Fundamental, CSIC, E-28006 Madrid, 
Spain
 
\noindent
$^{4}$ Department of Astronomy, Institut de Ci\`encies del Cosmos, Universitat de Barcelona
(UB-IEEC), Mart\'{\i} i Franqu\`es 1, E-08028 Barcelona, Spain 

\noindent
$^{5}$ Departamento  de Astrof\'{\i}sica y Ciencias de la Atm\'osfera,
Facultad de Ciencias F\'{\i}sicas, Universidad Complutense de Madrid, 
E-28040 Madrid, Spain

\noindent
$^{6}$ Isaac Newton Group of Telescopes, P.O. Box 321; E-38700 Santa 
Cruz de La Palma, Spain.

\noindent
$^{7}$ INAF-Osservatorio Astronomico di Padova, Vicolo dell'Osservatorio 5,
I-35122 Padova, Italy


\bigskip\medskip

{\bf

Type Ia supernovae are thought to occur as a white dwarf 
made of carbon and oxygen accretes sufficient mass to 
trigger a thermonuclear explosion$^{1}$. 
The accretion could occur slowly from an unevolved (main-sequence) 
or evolved (subgiant or giant) star$^{2,3}$, that being dubbed the 
single-degenerate channel, or rapidly as it breaks up a 
smaller orbiting white dwarf (the double-
degenerate channel)$^{3,4}$. 
Obviously, a companion will survive the
explosion only in the single-degenerate channel$^{5}$. 
Both channels might contribute to the production of type Ia 
supernovae$^{6,7}$ but their relative proportions still remain
a fundamental puzzle in astronomy. 
Previous searches for remnant companions have revealed one possible 
case for SN 1572$^{8,9}$, though that has been criticized$^{10}$. 
More recently, observations have restricted surviving companions to 
be small, main-sequence stars$^{11,12,13}$, ruling out giant 
companions, though still allowing the single-degenerate channel. 
Here we report the result of a search for surviving companions to 
the progenitor of SN 1006$^{14}$. 
None of the stars within 4' of the apparent site of the explosion is 
associated with the supernova remnant, so we can firmly exclude all 
giant and subgiant companions to the progenitor. 
Combined with the previous results, less than 20 per cent of 
type Iae occur through the single degenerate channel.

}

\bigskip

Together with SN 1572 (Tycho Brahe's supernova), 
SN 1604 (Kepler's supernova) and the recently identified SN 185, 
SN~1006 is one of only four known historical Galactic type Ia 
supernova events (SNeIa). 
It is also the only one whose Ia type has never been disputed. 
While a survey of the stars close to the centre of Tycho's 
supernova remnant (SNR) produced a likely candidate for the 
SN companion$^{8,9}$ and thus may be attributed 
to the single-degenerate channel, the absence of any 
ex-companion in the supernova remnant SNR 0509--67.5, in the 
Large Magellanic Cloud (LMC), down to very faint magnitudes, 
strongly suggests that the SN explosion there was produced by a 
double-degenerate system$^{15}$. 
Although the aforementioned {\it direct} searches have excluded, 
up to now, red-giant companions, there is some evidence from 
nearby spiral galaxies that a fraction of SNeIa may have had 
companions of this type$^{16}$. This hypothesis, however, is 
challenged by the absence of ultraviolet emission
that would be expected at the beginning of the SN outburst$^{17}$. 

The distance to the remnant of SN~1006 (2.18~$\pm$~0.08 kpc away, 
as determined from the expansion velocity and the proper motion 
of the ejecta$^{18}$), is much more precisely known than that of 
SN 1572 SNR (2.83$\pm$0.70 kpc).  
The interstellar extinction is also much smaller, 
which makes the distances to the stars less uncertain. To be a 
possible candidate, a star must first be at the correct distance and, 
depending on its spectral type and luminosity class, show some
spectral peculiarity, or an enhancement of the abundances of Fe-peak
elements such as that seen in star G of Tycho$^{9}$. 
Because of its relative vicinity compared with SN 1572 and the lower 
extinction ($A_{v} = 0.3$~mag, $vs.$ 2.0--2.4 mag in Tycho), 
going down to a red magnitude $m_R = 15$ allows us to include, 
up to the SN distance and beyond, all red giants, all subgiants, and 
also main-sequence stars down to an absolute magnitude 
$M_{R}\simeq +3.1$. 
Inspecting the 2MASS$^{19}$ photometric catalogue 
for possible unevolved companions (Supplementary Information), 
we found no main-sequence stars brighter that $m_R \simeq 16.4$, 
which brings the limit down to $M_{R} \simeq +4.5$, corresponding 
to $M_{V} \simeq +4.9$ (approximately equal to, or 
slightly less than, solar luminosity). 
Slightly evolved companions could be somewhat brigher: 
$m_R \simeq 16.0$ (just about twice as luminous as the Sun). 
Only in the case of SNR 0509--67.5 in the LMC$^{15}$ has a 
fainter limit has been reached in a direct search.

We derive the stellar atmosphere parameters of the sample stars 
(Fig.~1) using the high-resolution UVES spectroscopic data (Table~1).  
The spectra of four giants, and F- and G-type dwarfs of the sample 
are shown in Fig.~2, and in Supplementary Fig.~S2, respectively.
In Table~1 we also provide the radial velocities of sample stars 
measured from the UVES spectra, and distances determined from five 
photometric magnitudes, taking into account the stellar parameters
(Supplementary Information).
 In Fig.~3 we compare the abundances of Fe-peak elements in the stars 
of our sample with the Galactic trends$^{26}$, for F-, G- 
and K-type unevolved stars. Unlike in SN 1572, where star G shows an 
overabundance of Ni, [Ni/Fe]= 0.16 $\pm$ 0.04 (Here [A/B]$= \log
(N_A/N_B)-\log(N_A/N_B)_\odot$ for the number $N$ of atoms of elements
A and B, and subscript $\odot$ indicates the solar value), 
the stars in SN~1006 are all within the dispersion of the [Ni/Fe] 
Galactic trend. 
No enhancement is seen for any other element. 
In Supplementary Fig.~S3, we also 
show the Galactic trends of several $\alpha$-elements and we do not 
see any clear anomaly in these element abundances either.
  
None of the stars in the sample shows any significant rotation. 
High rotation speeds have been claimed to be a characteristic
of the surviving companions of SNeIa$^{10}$, based on the assumptions 
that, owing to tidal interaction and in spite of the angular mometum 
loss due to mass transfer, the rotation periods before explosion are 
equal to the (short) orbital periods, and that the radius of the star remains 
basically unchanged after the explosion$^{9}$. It has recently been 
shown$^{27}$, however, that the impact of the SNIa ejecta on the 
companion does indeed reduce those speeds by a large factor, 
which would make the rotation criterion irrelevant. 

Only four stars are at distances (marginally) compatible with that 
of the SNR. All of them are red giants: B16564 (G9-K0 III), 
B97341 (G9 III), B99810 (K1 III) and B93571 (K1 III) and none 
shows any spectroscopic peculiarity.  

Two-dimensional hydrodynamic simulations of the impact of the 
ejecta of a SNIa on a red-giant binary companion have been 
performed$^{28}$. More recently, the emission of supernova debris, 
arising from their impact with a similar companion, has been also 
considered$^{17}$. As for the effects of 
the explosion on the companion star, the results agree: most 
of the envelope is stripped away and what is left are the degenerate 
core and a very small fraction (amounting to a few per cent) of 
the original envelope. 
This hydrogen-rich envelope initially extends up 
to $\sim$350 R$_{\odot}$ and then contracts on a thermal timescale. 
The star should be evolving at constant luminosity  
($\sim$10$^{3}$ L$_{\odot}$) towards increasing effective 
temperatures for 10$^{5}$--10$^{6}$ yr. 
The same conclusion is reached when extrapolating to the red-giant 
case the results of simulations of the impact of SNIa 
ejecta on a main-sequence companion$^{29}$: the red giant 
should have been stripped off most of its hydrogen envelope. 
Nothing similar to any of the above four normal red giants
would be seen. On the other hand, the peculiar type of objects
that would result from the interaction of the SNIa ejecta with 
a red-giant companion would be luminous enough to have been 
seen at the distance of SN~1006 and within our magnitude limit.
A subgiant star similar to star G in Tycho would equally have been
seen, but the only three subgiants in the sample (B20292, B15360 and 
B14707) lie much closer than the SNR. 
We are thus left only with main-sequence 
stars, which lie at shorter distances than the SNR, which
have luminosities similar to, or lower than, 
the solar luminosity, and which are not predicted by any of the 
hydrodynamic simulations of the impact of SNIa ejecta with either 
a main-sequence star or a subgiant. In these hydrodynamic 
simulations$^{28}$, a main-sequence companion of 1 solar mass is 
puffed up and heated, reaching $\sim$5,000 times the solar 
luminosity; it is then predicted to contract and cool down on a 
thermal time scale. 
It has been found$^{30}$ that the return to luminosities 
of the order of those prior to the explosion could be faster 
because of the short cooling times of the outermost layers of the 
star, but, even so, in only $\sim$1000 yr the object does not have 
enough time to become dimmer than the Sun (Supplementary 
Information). 
 
Accordingly, SN~1006, the brightest event ever observed in  
our Galaxy, should have been produced either by mass accretion 
from an unevolved star, similar to, or less massive than, the Sun 
(with the above caveats), or by merging 
with another white dwarf. Adding this result to the evidence
from the other direct searches, the single-degenerate channel 
appears either to encompass only a clear minority of cases 
(20\% or less), or preferentially it involves main-sequence 
companions with masses more probably below that of the Sun.
              
{\bf References}

\begin{itemize}

\item[1] Nomoto, K., Saio, H., Kato, M., \& Hachisu, I. Thermal Stability
                  of White Dwarfs Accreting Hydrogen-rich Matter and 
		  Progenitors of Type Ia Supernovae. {\it Astrophys. J.}, 
                 {\bf 663}, 1269--1276 (2007)

\item[2] Patat, F., Chandra, P., Chevalier, R., et al. Detection of 
                 Circumstellar Material in a Normal Type Ia Supernova.
		 {\it Science}, {\bf 317}, 924--926 (2007) 

\item[3] Branch, D., Livio, M., Yungelson, L.R., Boffi, F., Baron, E. In 
                  search of the progenitors of Type Ia supernovae. 
		  {\it Publ. Astron. Soc. Pacif.}, {\bf 107}, 
		  1019--1029 (1995)

\item[4] Pakmor, R., Kromer, M., R{\"o}pke, F.~K., et al. Sub-luminous
                  type Ia supernovae from the mergers of equal-mass white 
		  dwarfs with mass ~0.9Msolar. {\it Nature}, 
		  {\bf  463}, 61--64 (2010)

\item[5] Ruiz--Lapuente, P. The quest for a supernova companion. 
                   {\it Science}, {\bf 276}, 1813--1814 (1997)

\item[6] Greggio, L. The rates of Type Ia supernovae -- II. Diversity of events
                  at low and high redshifts. {\it Month. Not. Royal Astron.
                  Soc.}, {\bf406}, 22--42 (2010)

\item[7] Brandt, T.D., {\it et al.} The ages of type Ia supernova 
          progenitors. {\it Astrophys. J.}, {\bf 140}, 804--816 (2010)

\item[8] Ruiz--Lapuente, P. {\it et al.} The binary progenitor of Tycho Brahe's
                        1572 supernova. {\it Nature}, {\bf 431}, 1069--1072
                       (2004)

\item[9] Gonz\'alez Hern\'andez, J.I., {\it et al.} 
                   The chemical abundances of Tycho G in supernova remnant 1572. 
		   {\it Astrophys. J.}, {\bf 691}, 1--15 (2009)

\item[10] Kerzendorf, W. {\it et al.} Subaru high resolution spectroscopy of 
                star G in the Tycho supernova remnant. {\it Astrophys. J}, 
                {\bf 701}, 1665--1672 (2009)

\item[11] Nugent, P.E., {\it et al.} Supernova SN 2011fe from an exploding 
                  carbon--oxygen white dwarf star. {\it Nature}, {\bf 480}, 
                  344--347 (2011)

\item[12] Li, W., {\it et al.} Exclusion of a luminous red giant as a companion
                  star to the progenitor of supernova SN 2011fe. {\it Nature}, 
                  {\bf 480}, 348--350 (2011)

\item[13] Edwards, Z.I., Pagnotta, A., Schaefer, B.E. The progenitor of the 
                  type Ia supernova that created SNR 0519--69.0 in the Large
                  Magellanic Cloud. {\it Astrophys. J.}, {\bf 747}, L19--L23
                  (2012)  

\item[14] Stephenson, F.R. SN~1006: the brightest supernova. {\it Astron. Geophys},
                    {\bf 51}, 5.27--5.32 (2010) 

\item[15] Schaefer, B.E., Pagnotta, A. An absence of ex--companion stars in the
                  type Ia supernova remnant SNR 0509--67.5. {\it Nature}, 
                  {\bf 481}, 164--166 (2012)

\item[16] Sternberg, A., {\it et al.} Circumstellar material in type Ia
          supernovae via sodium absorption features. {\it Science}, 
          {\bf 333}, 856--859 (2011)

\item[17] Kasen, D. Seeing the collision of a supernova with its companion 
               star. {\it Astrophys. J.}, {\bf 708}, 1025--1031 (2010)

\item[18] Winkler, P.F., Gupta, G., Long, K.S. The SN~1006 remnant: optical 
                   proper motions, deep imaging, distance, and brightness at 
                   maximum. {\it Astrophys. J.}, {\bf 585}, 324--335 (2003) 

\item[19] Cutri, R.~M., Skrutskie, M.~F., van Dyk, S., et al. 
                 2MASS All-Sky Catalog of Point Sources.
                 {\it VizieR Online Data Catalog}, {\bf 2246}, 0 (2003)


\item[20] Allen, G.E., Petre, R., Gotthelf, E.V. X--ray synchrotron emission 
                   from 10--100 TeV cosmic ray electrons in the supernova 
                   remnant SN~1006. {\it Astrophys. J.}, {bf 558}, 739--752 
                   (2001)

\item[21] Monet, D.G. The 526,289,881 objects in the USNO--A2.0 Catalogue. 
                  {\it Bull. Amer. Astron. Soc.}, {\bf 30}, 1427 (1998)

\item[22] Winkler, P.F., Long, K.S., Hamilton, A.J.S., Fesen, R.A. 
          Probing multiple sight lines through the SN 1006 remnant by
          ultraviolet absorption spectroscopy. {\it Astrophys. J.}, 
          {\bf 624}, 189--197 (2005) 

\item[23] Tabernero, H.~M., Montes, D., Gonz\'alez Hern\'andez, J.~I. 
          Chemically tagging the Hyades Supercluster.
          A homogeneous sample of F6-K4 kinematically-selected 
	  northern stars.
          {\it Astron. Astrophys.}, in press (2012).
	  Preprint at (http://arxiv.org/abs/astro-ph/1245.4879)
	  
\item[24] Sneden, C. PhD thesis. Univ. Texas, Austin (1973) 

\item[25] Kurucz, R.L. ATLAS89 Stellar Atmospheres Programs and 2 km s$^{-1}$ 
                  Grid. (CD--ROM, Smithsonian Astrophysical Observatory, 
                  Cambridge) (1993) 

\item[26] Neves, V., Santos, N.C., Sousa, S.G., Correia, A.C.M, Israelian, G.
                Chemical abundances of 451 stars from the HARPS GTO planet
                search program. Thin disk, thick disk, and planets 
                {\it Astron. Astrophys.}, {\bf 497}, 563--81 (2009)

\item[27] Pan, K.--C., Ricker, P., Taam, R. Impact of type Ia 
          supernova ejecta on the binary companions in the 
          single--degenerate scenario. {\it Astrophys. J.}, {\bf 750},  
          151 (2012) 

\item[28] Marietta, E., Burrows, A., Fryxell, B. Type Ia supernova explosions
              in binary systems: the impact on the secondary star and its
              consequences. 
              {\it Astroph. J. Suppl.}, {\bf 128}, 615--650 (2000)

\item[29] Pakmor, R., R\"opke, F.K., Weiss, A., Hillebrandt, W. The impact of 
               Type Ia supernovae on main sequence binary companions. 
               {\it Astron. Astrophys.}, {\bf 489}, 943--951 (2008)

\item[30] Podsiadlowski, P. On the evolution and appearance of a surviving
                       companion after a Type Ia supernova explosion. 
                       Preprint at (http://arxiv.org/abs/astro-ph/0303660) 
                       (2003)
 
\end{itemize}

\vfill\eject

\noindent{\bf Supplementary Information} is linked to the online
version of the paper at www.nature.com/nature. 

\noindent

\bigskip

\noindent{\bf Acknowledgements} 
This work was supported by the 
Spanish Ministerio de Ciencia e Innovaci\'on (MICINN), 
the Universidad Complutense de Madrid (UCM), and the 
Comunidad de Madrid.
This work is based on observations collected with the UVES 
spectrograph at the VLT/UT2 8.2-m Kueyen Telescope
(ESO run ID. 69.D-0397(A)) at the Paranal Observatory, Chile. 
We are grateful to the Cerro Paranal Observatory staff and to the
User Support Department of ESO for their help.

\noindent

\bigskip

\noindent{\bf Author Contributions} 
J.I.G.H. performed the chemical abundance analysis of the 
observed UVES spectra and determined the distances to the targets. 
J.I.G.H. and P.R.L. wrote the paper. 
P.R.L. was the Principal Investigator of the ESO proposal. 
R.C. and J.M. also participated in the ESO proposal. 
H.M.T. and D.M. derived the stellar parameters and created 
figures with the observed spectra. 
R.C. and P.R.L. contributed to the astrophysical interpretation 
of the results.
J.M. and L.R.B. collected the photometric data and created figures
with the field and the supernova remnant.
All the authors provided helpful comments and contributed to 
improve the text of the final version paper.   

\noindent 

\bigskip

\noindent{\bf Author information} 
Correspondence and requests should be addressed to J.I.G.H and P.R.L 
(jonay@iac.es; pilar@am.ub.es) 

\noindent 

\clearpage

\begin{table}

\begin{center}

{\bf Table 1 \textbar ~Parameters of the sample stars.} 

\bigskip

\begin{tabular}{llccrrc}
\\
\hline
\hline
{Name} & {$T_{\rm eff}$} & {$\log g$} & {$v_{\rm turb}$} 
& {[Fe/H]} &  {$v_{\rm rad}$} & {$d$} \\
\hline
 & [K] & [cgs] & [km/s] & [dex] & [km/s] & [kpc] \\
\hline
B09472 &  5853$\pm$75     & 4.52$\pm$0.17 & 1.211$\pm$0.105     &-0.04$\pm$0.06 &-20.74$\pm$2.14 & 1.23$\pm$0.53 \\
B11408 &  4677$\pm$68     & 2.91$\pm$0.23 & 2.136$\pm$0.082     &-0.29$\pm$0.05 &-123.54$\pm$1.48& 1.62$\pm$0.68 \\
B05723 &  5910$\pm$87     & 4.50$\pm$0.23 & 1.249$\pm$0.107     & 0.30$\pm$0.07 &-12.81$\pm$3.06 & 0.76$\pm$0.32 \\
B17720 &  5300$\pm$86     & 4.65$\pm$0.22 & 1.190$\pm$0.193     & 0.33$\pm$0.06 &-60.13$\pm$1.30 & 0.58$\pm$0.25 \\
B16564 &  4845$\pm$39     & 3.13$\pm$0.16 & 1.607$\pm$0.049     &-0.37$\pm$0.03 &-116.07$\pm$1.86& 3.03$\pm$1.27 \\
B97338 &  5707$\pm$42     & 4.21$\pm$0.12 & 1.292$\pm$0.054     & 0.01$\pm$0.04 &-5.53$\pm$2.25  & 1.05$\pm$0.45 \\
B05518 &  6327$\pm$63     & 4.85$\pm$0.13 & 1.386$\pm$0.090     & 0.14$\pm$0.04 &-4.38$\pm$1.92  & 0.50$\pm$0.21 \\
B20292 &  5196$\pm$33     & 3.67$\pm$0.09 & 1.372$\pm$0.051     &-0.66$\pm$0.03 & 2.92$\pm$2.52  & 1.13$\pm$0.48 \\
B14130 &  6177$\pm$44     & 4.61$\pm$0.09 & 1.528$\pm$0.052     & 0.31$\pm$0.03 &-21.26$\pm$2.09 & 0.89$\pm$0.38 \\
B03395 &  5517$\pm$49     & 4.36$\pm$0.13 & 1.165$\pm$0.074     & 0.22$\pm$0.04 &-56.18$\pm$5.21 & 0.95$\pm$0.41 \\    
B97341 &  4881$\pm$51     & 2.98$\pm$0.18 & 1.641$\pm$0.057     &-0.16$\pm$0.04 &$-$41.76$\pm$1.96 & 2.48$\pm$1.04 \\
B99810 &  4658$\pm$39     & 2.51$\pm$0.15 & 1.782$\pm$0.039     &-0.72$\pm$0.03 & 33.53$\pm$1.50 & 2.42$\pm$1.02 \\
B15360 &  4960$\pm$70     & 3.40$\pm$0.21 & 1.886$\pm$0.086     & 0.07$\pm$0.05 &-37.03$\pm$1.61 & 1.29$\pm$0.54 \\
B93571 &  4579$\pm$58     & 2.51$\pm$0.22 & 2.182$\pm$0.066     &-0.21$\pm$0.05 &-100.33$\pm$1.58& 2.39$\pm$1.01 \\
B18024 &  6083$\pm$58     & 4.44$\pm$0.13 & 1.545$\pm$0.095     &-0.36$\pm$0.04 & 22.61$\pm$2.35 & 0.60$\pm$0.26 \\
B08277 &  5693$\pm$34     & 4.39$\pm$0.09 & 1.123$\pm$0.044     & 0.17$\pm$0.03 &-49.50$\pm$2.08 & 0.45$\pm$0.19 \\
B24215 &  5729$\pm$45     & 4.44$\pm$0.12 & 1.189$\pm$0.058     & 0.16$\pm$0.04 & 3.02$\pm$2.39  & 0.95$\pm$0.41 \\
B14707 &  5065$\pm$47     & 3.36$\pm$0.15 & 1.270$\pm$0.060     & 0.16$\pm$0.04 & 22.14$\pm$1.83 & 1.37$\pm$0.58 \\
B90474 &  5051$\pm$38     & 3.05$\pm$0.12 & 1.874$\pm$0.046     &-0.29$\pm$0.03 &-98.10$\pm$1.95 & 4.78$\pm$2.00 \\
B90102 &  5650$\pm$33     & 4.21$\pm$0.10 & 1.076$\pm$0.049     &-0.11$\pm$0.03 & 85.57$\pm$1.16 & 0.91$\pm$0.39 \\
B10074 &  5601$\pm$63     & 4.52$\pm$0.14 & 1.036$\pm$0.102     & 0.05$\pm$0.05 &-30.26$\pm$1.38 & 0.71$\pm$0.30 \\
B95979 &  6776$\pm$94     & 4.19$\pm$0.16 & 1.832$\pm$0.131     & 0.02$\pm$0.06 &-32.94$\pm$6.81 & 0.94$\pm$0.40 \\
B09749 &  6163$\pm$65     & 4.53$\pm$0.14 & 1.362$\pm$0.082     & 0.14$\pm$0.05 &-13.15$\pm$2.60 & 0.40$\pm$0.17 \\
B26090 &  6101$\pm$39     & 4.58$\pm$0.09 & 1.238$\pm$0.053     & 0.15$\pm$0.03 &-48.08$\pm$2.52 & 0.37$\pm$0.16 \\
\hline
\end{tabular}
\end{center}

\label{tpar}

\noindent{}{{\bf Table 1 \textbar ~Parameters of the sample stars.} 
We provide the stellar atmosphere parameters (effective temperature,
{$T_{\rm eff}$}, surface gravity, {$\log g$}, and microturbulent
velocity, {$v_{\rm turb}$}), metallicities, [Fe/H], 
radial velocities, {$v_{\rm rad}$}, and distances, $d$, of 
the sample stars, together with 1-$\sigma$ uncertainties.
The stellar parameters and metallicities were 
obtained from Fe~{\scshape I} and Fe~{\scshape II} excitation 
and ionization equilibria, using the {\scshape StePar}$^{23}$ code,
which makes use of the MOOG$^{24}$ code and ATLAS9 model 
atmospheres$^{25}$. 
Our analysis has been
performed assuming local thermodynamic equilibrium (LTE).
SN~1006 is located about 500 pc above the Galactic plane. 
Unlike the case of SN 1572, which lies close to the Galactic plane, 
the radial velocities of the stars along the line of sight do not 
follow a regular pattern (this can be seen in the sixth column), 
since most of them belong either to the thick disc 
or to the halo of the Galaxy (see the dispersion in metallicities 
in the fifth column). Two of the observed stars, B98824 and B21185,
are spectroscopic binaries and were discarded as possible companion
stars of the progenitor of SN 1006 (Supplemetary information).}

\end{table}

\clearpage

\begin{figure}[ht!]
\centering
\includegraphics[width=0.8\textwidth,angle=0]{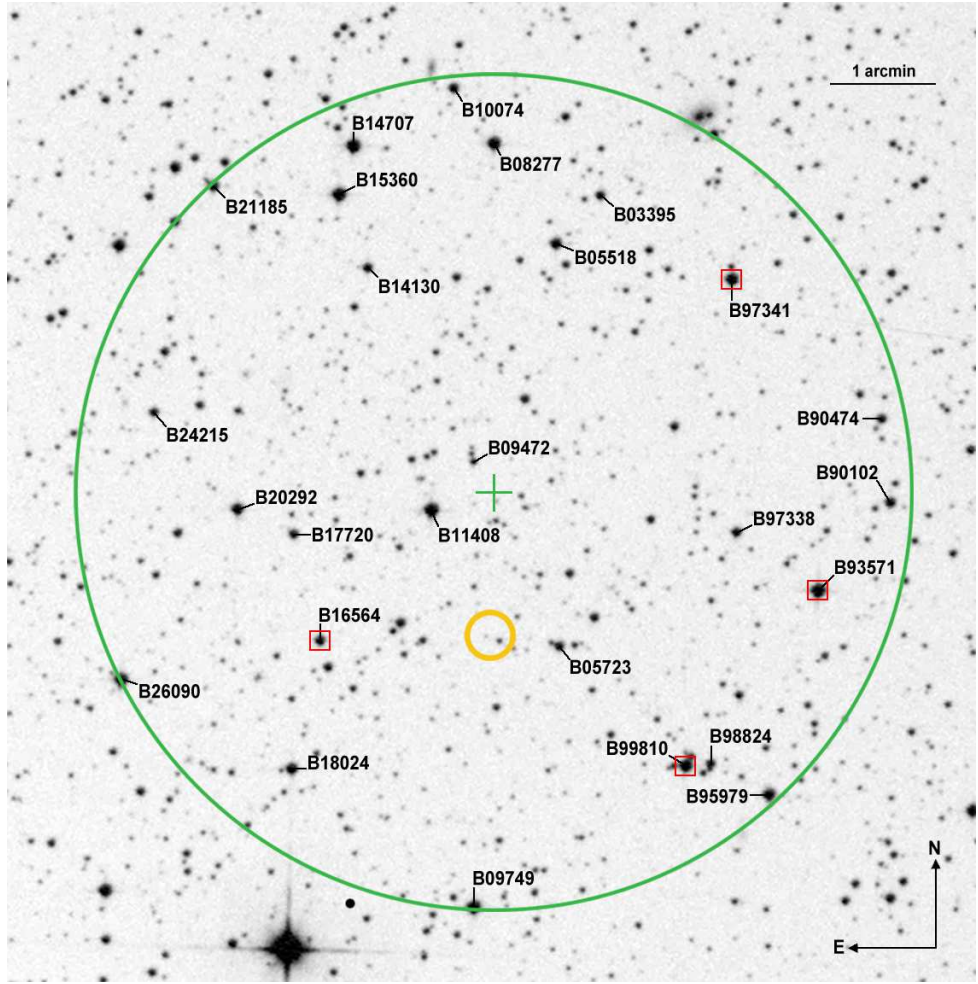}
\caption{{\bf \textbar ~DSS2 R-band image of the SN 1006 field.} 
The positions and names 
of the 26 stars included in the spectroscopic survey (selected from 
the USNO-A2.0 Catalogue$^{21}$) are given. 
The centre of our search is the geometrical centre of the quite
symmetrical X-ray emission of the SNR$^{20}$ (green cross),  
at RA = 15 $^{h}$ 2 $^{m}$ 55 $^{s}$, Dec = $-$41 $^\circ$ 
55$^\prime$ 12$^{\prime\prime}$. 
Also shown are the boundary of the surveyed region (large green
circle) and the geometrical centre of the H-alpha emission$^{18}$ 
(small yellow circle).
Giant stars are marked by red squares. 
See also Supplementary Fig.~S1, for a full view of the SNR. 
Another centre has been proposed more recently$^{22}$, 
based on the distribution of the ejecta along the line of sight. 
It is, however, still located within our surveyed area. For a star at 
a distance $d \simeq 2.2$ kpc and a velocity perpendicular to the 
line of sight of $\sim$100 km s$^{-1}$ 
(roughly the orbital velocity before the explosion), 
the angular displacement on the sky in 1000 years would be 
10$^{\prime\prime}$ only. However, given the asymmetry of the SNR, 
and also that in core-collapse SNe the distance between the compact
object and the X-ray centroid of the SNR can be 15\% of the radius 
of the SNR or more, we adopted a much wider radius for the search:
4$^{\prime}$. 
That amounts to 27\% of the radius of the SNR, which is
15$^{\prime}$ (Supplementary Fig.~S1).  
Positions, magnitudes and angular distances to the centre of the
survey are given in Supplementary Table~S1.}
\label{fsky}
\end{figure}

\clearpage

\begin{figure}[ht!]
\centering
\includegraphics[width=0.8\textwidth,angle=0]{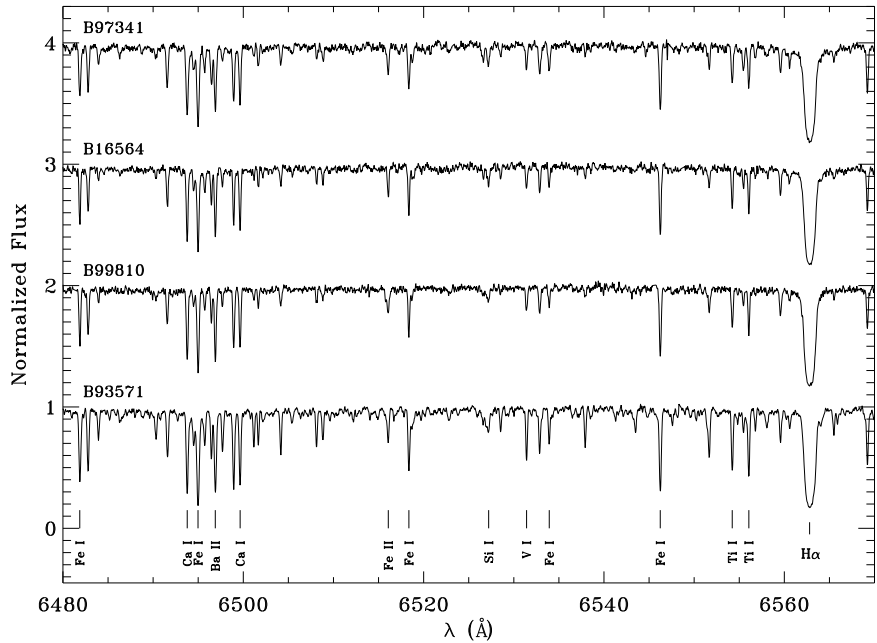}
\nopagebreak[4]
\label{fspec}
\caption{{\bf \textbar ~Observed UVES spectra of the candidate giant stars.} 
The spectra are labelled by their names in Table~1 and sorted 
in decreasing effective temperature
order from top to bottom. 
These high-resolution spectra were 
obtained at the 8.2-m Kueyen VLT (UT2) telescope equipped 
with UVES at the Cerro Paranal Observatory in Chile. 
They were obtained on 13, 14, 25, 28--30  
April 2002 and on 1 May 2002, covering the spectral regions 
3295--5595~{\AA}, 5680--6645~{\AA}, 6705--8515~{\AA}, and 
8675--10420~{\AA} at resolving power $\lambda/\Delta\lambda 
\approx 43\,000$. The S/N of the spectra is in the range 50--170 
and typically 80. The data were reduced in a standard manner, and 
later normalized within IRAF, using low-order polynomial fits 
to the observed continuum.}
\end{figure}

\clearpage

\begin{figure}[ht!]
\centering
\includegraphics[height=0.8\textwidth,angle=90]{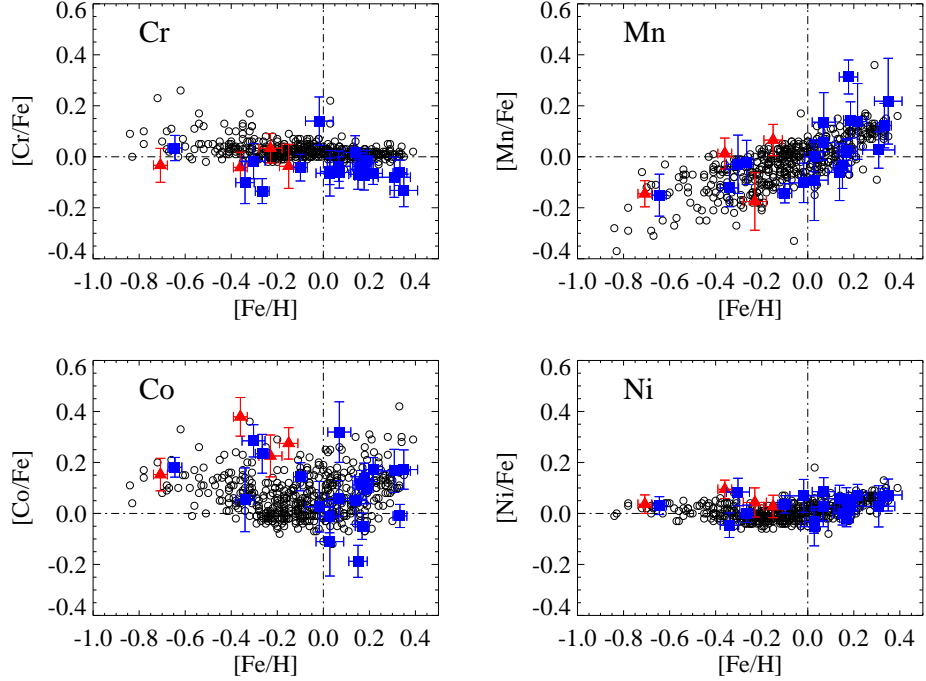}
\nopagebreak[4]
\label{fabun}
\caption{{\bf \textbar ~Stellar abundance ratios [X/Fe] of several Fe-peak elements.}
The chemical abundances have been derived using the EW technique 
for all of the elements. We performed a differential analysis
on a line-by-line basis, using the solar UVES spectrum of the Moon 
as reference (see the Supplementary Section 3 for more details). 
The abundances of several Fe-peak elements are listed in 
Supplementary Table~S4b relative to iron are compared with 
the Galactic trends of these elements in the relevant range of 
metallicities$^{26}$. 
Red triangles correspond to the four giant stars whose distances 
are marginally compatible with that of the remnant of SN~1006. Blue 
squares, to the rest of the stars in the sample.
The error bars are the 1-$\sigma$ uncertainties associated with the 
dispersion of the measurements from different spectral features.
}
\end{figure}

\clearpage

\noindent
{\Large \bf SUPPLEMENTARY INFORMATION}

\smallskip

\noindent
{\bf 1.  Luminosities of the surviving companions}

\noindent

As stated in the main text, hydrodynamic simulations$^{28}$ of the
interaction of the SNIa ejecta with either a main-sequence or a
subgiant companion predict a significant injection of energy into the
outer layers of the companion star that remain gravitationally bound
after the impact.  
That results in a large excess in radius and in luminosity of the
companion, as compared with the values corresponding to its mass and
evolutionary stage, in hydrostatic and thermal equilibrium (the
estimate is as high as $\sim$5,000 L$_{\odot}$). 
Simulations$^{29}$ of the collision of the material ejected by a
SNIa with a main-sequence companion are entirely consistent with 
earlier ones$^{28}$, the differences in the amount of mass stripped 
from the companion being due only to the more compact initial stellar
models adopted for the simulations. 
The question still remains of how fast thermal equilibrium is 
restored 
(hydrostatic equilibrium is recovered very shortly after the 
collision with the SNIa ejecta). 
This issue has been explored for the case of main-sequence and 
subgiant companions$^{30}$. 
Contrary to the assumption$^{28}$ that the governing time scale
for the evolution in luminosity
should be the thermal time scale of the whole stellar envelope, it
is found that the luminosity can decrease much faster, due to the 
short cooling times of the outermost layers, so that in a much 
reduced time it could return (roughly) to its pre-explosion value.

The luminosity evolution depends on the amount of mass stripped, 
on the energy injected into the bound layers and on the cooling times
of the outermost part of the stellar envelope. For values matching the
results of the hydrodynamic simulations$^{28}$ (and for a subgiant
companion), the luminosities predicted$^{30}$ for a time $\sim$1000
yr after the SNIa explosion are within $\sim$10--25$L_{\odot}$. Thus,
even allowing for a wide margin of uncertainty, nothing below 
solar luminosity is to be expected.

A possibility has been suggested, within the SD channel, involving a 
low luminosity companion of the exploding white dwarf 
at the time of the SNIa$^{31,32}$. The idea is that mass transfer 
from the companion spins up the accreting white dwarf, making the 
critical mass for explosive carbon ignition, $M_{\rm crit}$, higher 
than the Chandrasekhar mass ($M_{\rm Ch}$ $\simeq$ 1.4 $M_{\odot}$). 
If mass transfer 
ceases when the mass of the white dwarf, $M$, is still lower than
the $M_{\rm crit}$ corresponding to its spin rate (but higher than 
$M_{\rm Ch}$), there should be some
time interval $\tau$ between the end of mass accretion and the SNIa,
required for spinning down until $M_{\rm crit} = M$. 
It is then speculated$^{32}$
that $\tau$ could be long enough for the companion (if it were a
red giant or even a subgiant star) to have become a white dwarf. 
It must
be noted that solid-body rotation only very slightly increases 
$M_{\rm crit}$ above $M_{\rm Ch}$ ($M_{\rm crit} = 1.47\ M_{\odot}$). 
Only differential
rotation can significantly increase $M_{\rm crit}$$^{33,34}$, but, in 
any case, the upper limit to the time scale for angular momentum 
redistribution and loss appears to be $\sim$10$^{6}$ yr only$^{33}$. 
That would be enough, 
if the mass donor were a red-giant star and mass transfer stopped 
owing to contraction of the envelope when its mass becomes too low 
to sustain the red-giant structure$^{31}$, to have a companion with 
a radius much smaller than the Roche lobe radius, but it would not 
be much less luminous than during the previous stage yet. 
To assume arbitrarily long time scales for 
angular momentum redistribution and loss (thus allowing the 
companion star to become a white dwarf) is not supported by current 
models$^{35,36}$.      


\bigskip

\begin{figure}[ht!]
\centering
\includegraphics[width=0.8\textwidth,angle=0]{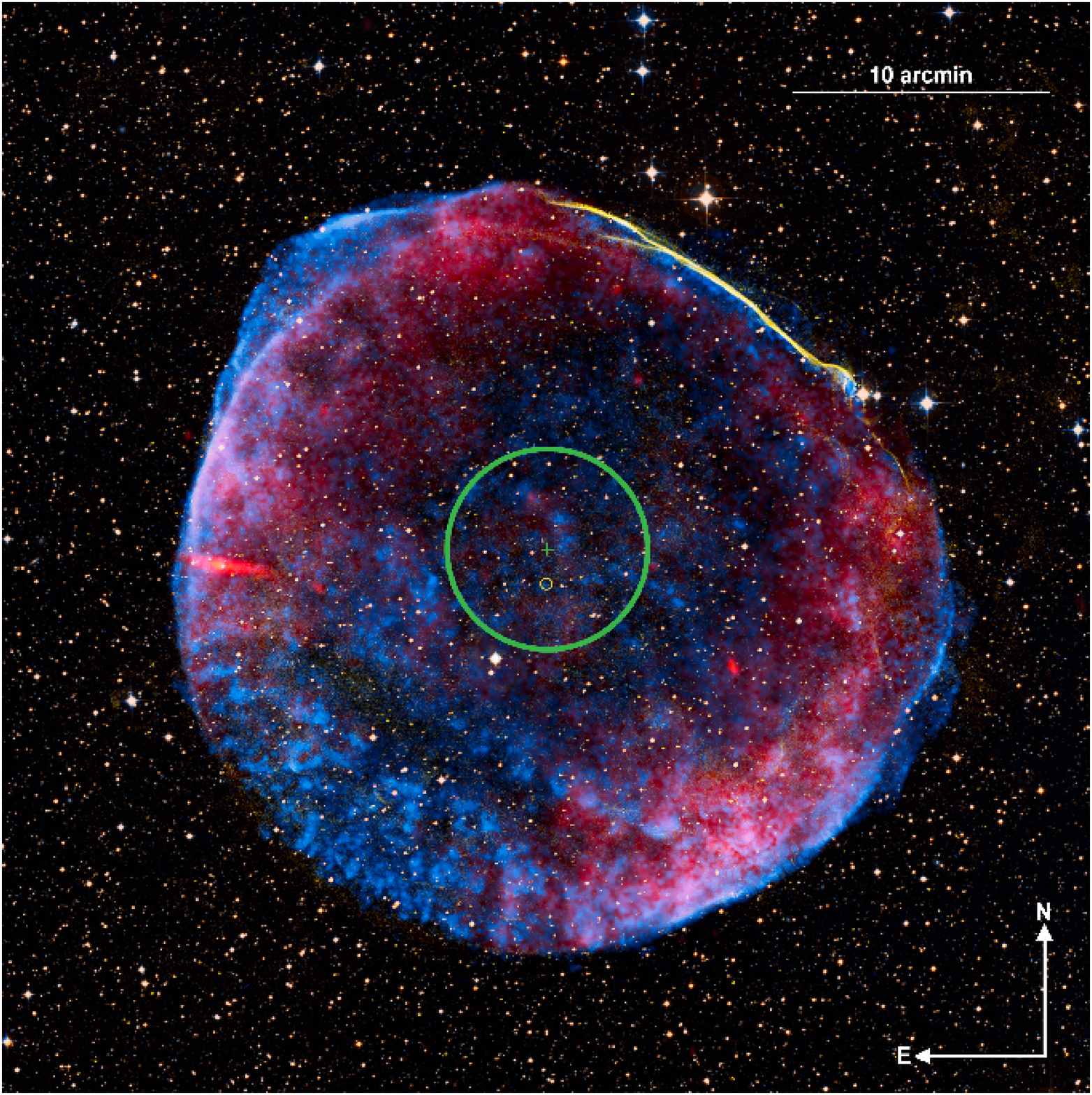}
\bigskip
\noindent

\begin{minipage}[b]{0.9\textwidth}
{\bf Figure S1 \textbar ~The remnant of the SN1006} 
The surveyed area is indicated by the large 
green circle. The centre of the survey (the centroid of the X-ray 
emission) is marked with a green cross, and that of the H$\alpha$ 
emission, by the small yellow circle. 
This is a composite image of the SN 1006 supernova remnant, which is
located about 7000 light years from Earth. Shown here are X-ray data
from NASA's Chandra X-ray Observatory (blue), optical data from the
University of Michigan's 0.9 metre Curtis Schmidt telescope at the
NSF's Cerro Tololo Inter-American Observatory (CTIO; yellow) and the
Digitized Sky Survey (orange and light blue), plus radio data from 
the NRAO's Very Large Array and Green Bank Telescope (VLA/GBT; red). 
Image credit: X-ray: NASA/CXC/Rutgers/G.Cassam-Chena\"i, J.Hughes et
al.; Radio: NRAO/AUI/NSF/GBT/VLA/Dyer, Maddalena \& Cornwell; Optical:
Middlebury College/F.Winkler, NOAO/AURA/NSF/CTIO Schmidt \& DSS 
\end{minipage}

\label{frem}
\end{figure}

\clearpage

\smallskip

\noindent
{\bf 2.  Distance determinations of the sample stars}

\noindent

We have estimated the distances of the sample stars from the 
photometric magnitudes in five
different filters: Johnson $m_B$ and $m_R$, and $m_J$, $m_H$ and 
$m_K$ from the 2MASS$^{19}$
catalogue (see Supplementary Table~S2). 
The masses of the stars were estimated using a code$^{37,38,39}$ 
kindly provided by C. Allende-Prieto, according to their stellar 
parameters: effective temperature, {$T_{\rm eff}$}, and surface
gravity, $\log g$, and their metallicity, [Fe/H], and assuming for 
all stars the uncertainties of 100~K, 0.3~dex and 0.1~dex in 
{$T_{\rm eff}$}, $\log g$ and [Fe/H], respectively.  
The calculations are based on solar-scaled theoretical 
isochrones$^{40}$.
The radii were then derived from the surface gravity and mass. 
This radius, together with the spectroscopic estimate of the
effective temperature, provides the 
intrinsic bolometric luminosity in the ranges
$0.3 < L_\star/L_\odot < 1.4$ for dwarfs,
$4 < L_\star/L_\odot < 14$ for subgiants, and 
$19 < L_\star/L_\odot < 69$ for giants.

In the Supplementary Table~S3, we show the distance 
determinations for the different filters, using the bolometric 
corrections for models without overshooting$^{41}$, 
for our preferred value of the colour excess, $E(B-V)=0.096$ mag, 
and other sets of the relevant parameters. 
We compute the magnitude corrected for extinction in each
filter as $m_{V,0}=m_V-A_V$, where $A_V = 3.12\, E(B-V)$ is the
extinction in the Johnson $V$ filter.  We adopt the following values
for other filters$^{42}$:$A_B/A_V=1.31$ $A_R/A_V=0.84$,
$A_J/A_V=0.32$, $A_H/A_V=0.27$ and $A_K/A_V=0.21$. We adopt errors
in $m_B$ and $m_R$ of 0.03.
The error in the distance from each filter takes into account
uncertainties of 100~K and 0.3~dex in the stellar parameters, and 
uncertainties of 0.2~$M_\odot$ and 0.05 in mass and reddening, 
respectively, plus the individual errors of the photometric 
magnitudes. 
The final average distance of each star is given in Table~1, and 
Supplementary Tables S2 and S3 give the mean distances weighted 
by the individual uncertainties in different filters. 
We note that distance determinations from different filters are 
consistent and the final errors in the distances mostly reflect 
the systematic uncertainties.

The possibility that the proposed companion, Tycho G, of the SNIa 
SN 1572 had lost part of its envelope, $\Delta M=-0.2$ M$_\odot$, 
due to the impact of the supernova ejecta$^{28,29,30}$, and that that
would affect its derived distance, has been discussed$^{9}$. 
It was concluded that the lower mass does not produce a
significant change in the derived average distance, assuming that 
the subgiant companion star is able to almost completely 
recover thermal equilibrium between $\sim 10^2$ and $\sim 10^3$ yr 
after the white dwarf explosion$^{30}$.
We have done similar calculations and checked that this statement 
also applies to the sample stars in the SN~1006 field. 

Finally, we inspected the 2MASS$^{19}$ catalogue to search for any 
possible main-sequence candidate to be the companion of SN~1006.
For that, we extracted the 2MASS JHK magnitudes of the stars within 
the 4~arcmin radius around the centre of SN~1006 and computed the 
effective temperature from the IRFM$^{43}$ assuming that all stars in the field
are solar-metallicity ([Fe/H]$=0$), main-sequence ($\log g=4.5$) 
stars.
We determined their masses and luminosities as previously indicated 
and saw that there are only roughly 15 stars with magnitudes 
$m_R$ in the range 16.3--17.5 ($m_K$ in the range 15.1--15.7)  
and $T_{\rm eff}$ in the range 5250--6450~K. 
The same exercise has been repeated for slightly evolved stars,  
with $\log g=4$. 
In this case, there are just 3 stars with magnitudes $m_R$ in 
the range 16.0--16.3 ($m_K$ in the range 14.8--15.1) and 
$T_{\rm eff}$ in the range 5350--5500~K. 
This upper-limit in the apparent magnitude corresponds to                                       
an absolute magnitude of $M_R=+4.1$. The case for 
$\log g=3.5$ produces stars at distances far beyond that of 
SN~1006. Therefore, as stated in the main text, any possible
main-sequence companion at the distance of the SNR
should be about as luminous as the Sun or less. 
Slightly evolved companions should have roughly twice the 
solar luminosity at most. 

\bigskip

\noindent
{\bf 3. Determination of stellar parameters and element abundances 
of the sample stars}

The stellar atmosphere parameters of the sample stars were 
obtained from the Fe~{\scshape I} and Fe~{\scshape II} excitation 
and ionization equilibria. 
We derive the equivalent widths (EWs) of the
Fe~{\scshape I} and Fe~{\scshape II} lines with the 
{\scshape ARES} code$^{44}$ using a line list with 
263 Fe~{\scshape I} and 36 Fe~{\scshape II} lines$^{45}$. 
The parameters were computed using the {\scshape StePar} code$^{23}$. 
This code employs the 2002 version of the MOOG code$^{24}$, and a 
grid of Kurucz ATLAS9 plane--parallel model atmospheres$^{25}$. 
Our analysis has been performed assuming local thermodynamic 
equilibrium (LTE).
The parameters are shown in columns 2--4 of Table 1. 

The chemical abundances have been derived by computing the EWs 
of spectral lines using the ARES code$^{44}$ for all  
the elements. Once the EWs are measured, we use the LTE 
MOOG code$^{24}$ to compute the chemical abundance provided by each 
spectral line, using the appropriate ATLAS9 model atmosphere$^{25}$ 
for each star. 
We determine the mean abundance of each element relative to its 
solar abundance (using the UVES spectrum of the Moon) by 
computing the line-by-line mean difference. In 
Supplementary Tables S4a and S4b, we provide the average abundances 
of each element, together with the errors associated with the 
dispersion of the measurements from different spectral features.
In Fig.~3 and Supplementary Fig.~S3 we compare the abundance ratios 
[X/Fe] with the Galactic trends$^{26}$ and we do not see any clear 
signature of enhancement in the abundances of $\alpha$-elements 
or Fe-peak elements in any of the sample stars. 
The high scatter in the abundance ratios of 
some elements is related to the small available number of lines 
in the spectra.  

\bigskip

{\bf Supplementary References}

\begin{itemize}

\item[31]  Justham, S. Single degenerate type Ia supernovae without hydrogen
                 contamination. {\it Astrophys. J. (Letters)}, {\bf 730}, 
                 L34--L38 (2011)

\item[32]  Di Stefano, R., Voss, R., Claeys, J.S.W. Spin--up/spin--down 
                 models for Type Ia supernovae. {\it Astrophys. J.}, 
                 {\bf 738}, L1--L4 (2011) 

\item[33]  Yoon, S., Langer, N. Presupernova evolution of accreting white dwarfs
                          with rotation. 
                          {\it Astron. Astrophys.}, {\bf 419}, 623--644
                          (2004)

\item[34]  Yoon, S., Langer, N. On the evolution of rapidly rotating massive
                          white dwarfs towards supernovae or collapses. 
                          {\it Astron. Astrophys.}, {\bf 435}, 967--985
                          (2005)

\item[35]  Saio, H., Nomoto, K. Off--center carbon ignition in rapidly 
                          rotating, accreting carbon--oxygen white dwarfs.   
                          {\it Astrophys. J.}, {\bf 615}, 444--440 
                          (2004)

\item[36]  Piro, A.L.     The internal shear of Type Ia supernova progenitors
                          during accretion and simmering.
                          {\it Astrophys. J.}, {\bf 679}, 616--625
                          (2008) 

\item[37]  Allende-Prieto, C., Asplund, M., Fabiani Bendicho, P. 
                          S4N: A spectroscopic survey of stars in the solar
                          neighborhood. The Nearest 15 pc.         
                          {\it Astron. Astrophys.}, {\bf 423}, 1109--1117 (2004)

\item[38]  Reddy, B.~E., Lambert, D.~L., Allende Prieto, C. Elemental 
                         abundance survey of the Galactic thick disc. 
                         {Monthly Not. Royal Astron. Soc.}, {\bf 367}, 
                         1329--1366 (2006)
 
\item[39]  Ram{\'{\i}}rez, I., Allende Prieto, C., Lambert, D.~L. 
                           Oxygen abundances in nearby stars. Clues to the 
                           formation 
			   and evolution of the Galactic disk. 
			   {\it Astron. Astrophys.}, {\bf 465}, 271--289 (2007)

\item[40]  Bertelli, G., Bressan, A., Chiosi, C., Fagotto, F., Nasi, E.
                          Theoretical isochrones from models with new radiative
			  opacities.
			  {\it Astron. Astrophys. Suppl.}, {\bf 106}, 275--302 
                          (1994) 

\item[41]  Bessell, M. S., Castelli, F., Plez, B.
                           Model atmospheres broad-band colors, 
			   bolometric corrections and temperature 
			   calibrations for O - M stars. 
 			   {\it Astron. Astrophys.}, {\bf 333}, 231--250 (1998)

\item[42]  Schaifers, K., {\it et al.} Astronomy and Astrophysics. C: 
                           Interstellar Matter, Galaxy, Universe, in 
                           Landolt--B\"orstein: Numerical Data and functional
                           Relationships in Science and Technology. New Series
                           (Springer--Verlag, Berlin) (1982) 

\item[43]  Gonz{\'a}lez Hern{\'a}ndez, J.~I., \& Bonifacio, P. A new 
                          implementation of the infrared flux method using 
			  the 2MASS catalogue. {\it Astron. Astrophys.},
			  {\bf 497}, 497--509 (2009) 

\item[44] Sousa, S.G., Santos, N.C., Israelian, G., Mayor, M., 
          Monteiro, M.J.P.F.G. A new code for automatic determination 
	  of equivalent widths: Automatic Routine for line Equivalent 
	  widths in stellar Spectra (ARES). 
	  {\it Astron. Astrophys.}, {\bf 469}, 783--791 (2007)  

\item[45] Sousa, S.G. {\it et al.} Spectroscopic parameters for 451 stars in 
          the HARPS GTO planet search program. Stellar [Fe/H] and the
          frequency of exo--Neptunes. {\it Astron. Astrophys.}, {\bf 487}, 
          373--381 (2008)  

\item[46]  Gray, D. F. The observation and analysis of stellar 
                       photospheres.
                       {\it Camb.~Astrophys.~Ser.}, {\bf Vol.~20} (1992)

\end{itemize}

\clearpage

\begin{figure}[ht!]
\centering
\includegraphics[width=0.7\textwidth,angle=0]{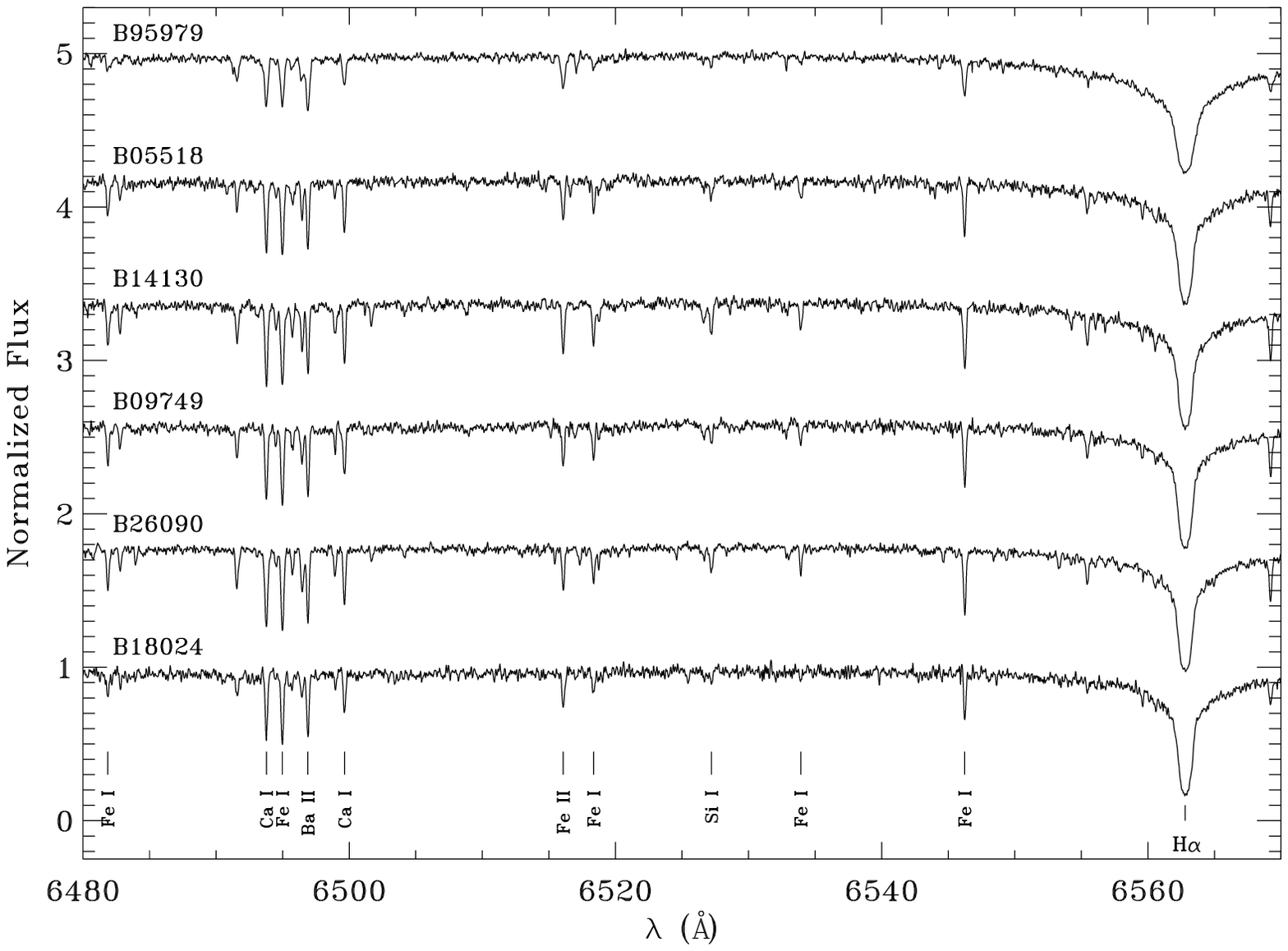}
\includegraphics[width=0.7\textwidth,angle=0]{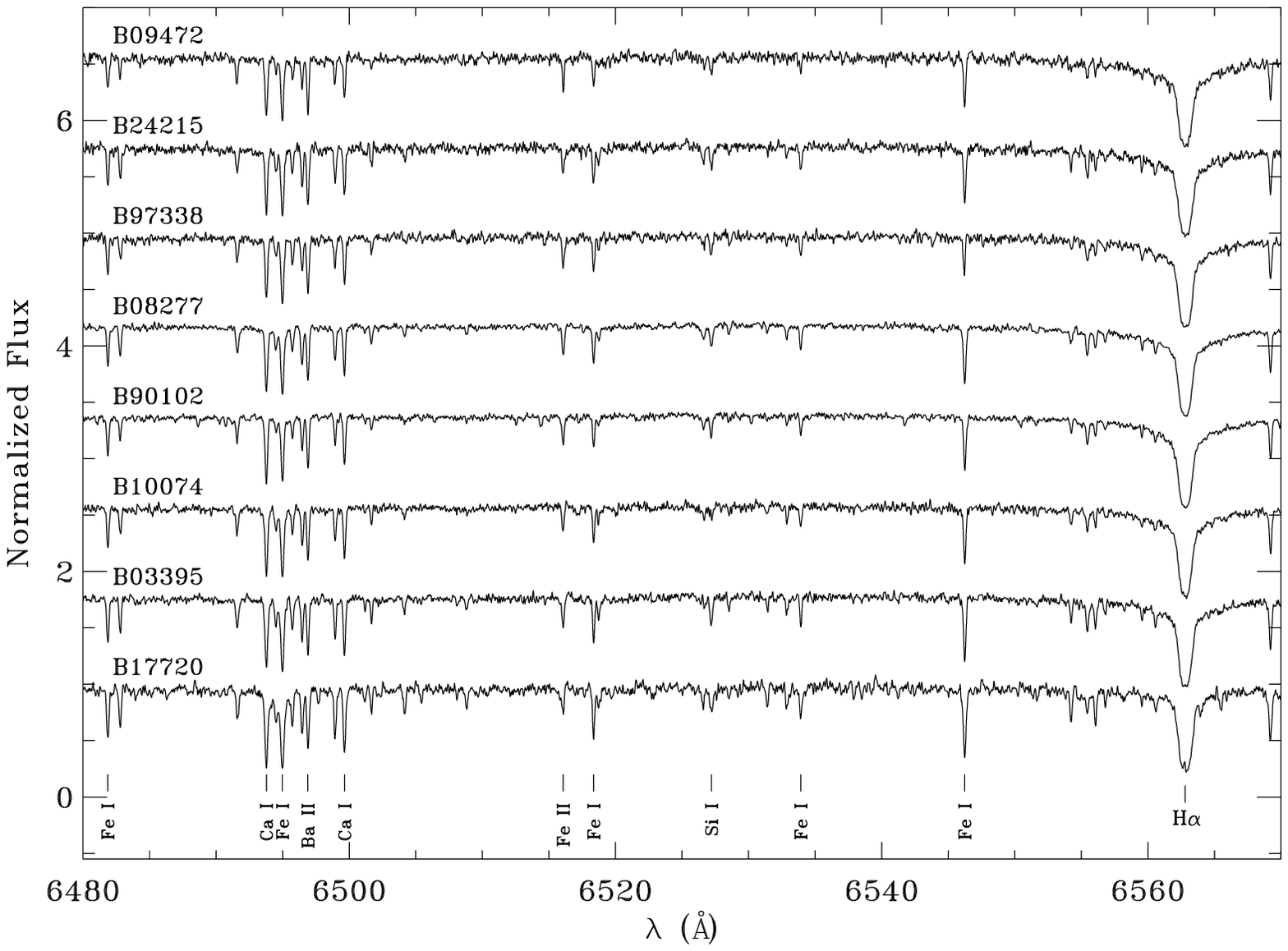}
\nopagebreak[4]

\bigskip
\noindent

\begin{minipage}[b]{0.9\textwidth}
{\bf Figure S2 \textbar ~Observed UVES spectra of the candidate dwarf stars.} 
Spectra of the F-type (top panel) and G-type (bottom panel) sample 
stars, labelled by their names in Table~1, and sorted in decreasing 
effective temperature order from top to bottom.
These high-resolution spectra were 
obtained at the 8.2-m Kueyen VLT (UT2) telescope equipped 
with UVES at the Cerro Paranal Observatory in Chile. 
They were obtained on 13, 14, 25, 28--30  
April 2002 and on 1 May 2002, covering the spectral regions 
3295--5595~{\AA}, 5680--6645~{\AA}, 6705--8515~{\AA}, and 
8675--10420~{\AA} at resolving power $\lambda/\Delta\lambda 
\approx 43\,000$. The S/N of the spectra is in the range 50--170 
and typically 80. The data were reduced in a standard manner, and 
later normalized within IRAF, using low-order polynomial fits 
to the observed continuum.
\end{minipage}

\label{fspecb}
\end{figure}

\clearpage

\begin{figure}[ht!]
\centering
\includegraphics[height=0.8\textwidth,angle=90]{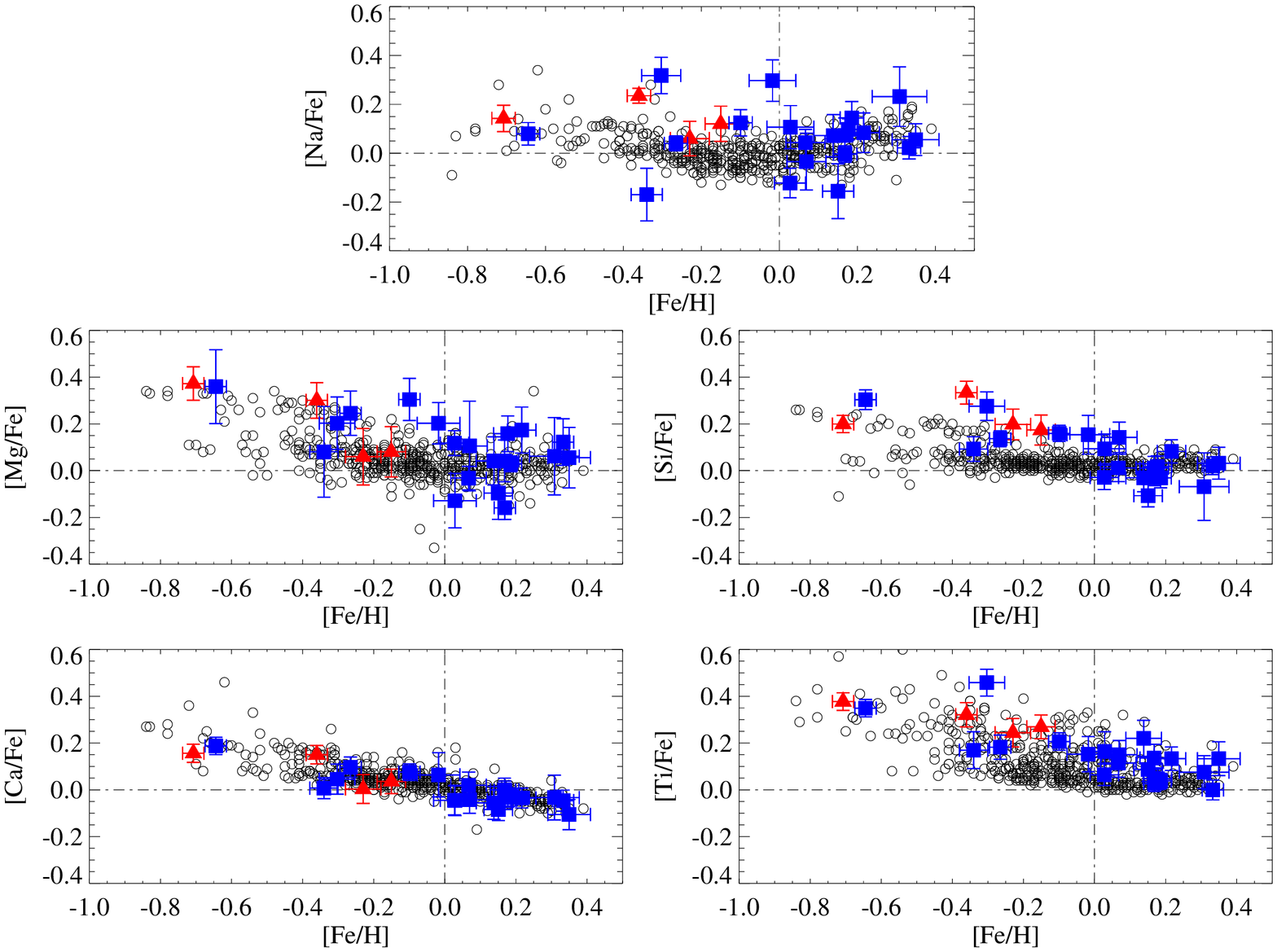}
\nopagebreak[4]

\bigskip
\noindent

\begin{minipage}[b]{0.9\textwidth}
{\bf Figure S3 \textbar ~Stellar abundances of several elements.}
These element abundances are listed in Supplementary Table~S4a
relative to iron are compared with the Galactic trends of these 
elements in the relevant range of metallicities$^{26}$. 
Red triangles correspond to the four giant stars whose distances 
are marginally compatible with that of the remnant of SN~1006. Blue 
squares, to the rest of stars in the sample.
\end{minipage}

\label{fabunb}
\end{figure}

\clearpage

\begin{table}

\small

{\bf Table S1 \textbar ~Astronomical positions, photometric magnitudes
$m_{R}$ and $m_{B}$, and angular distance to the center of the survey,
of the sample stars}

\smallskip
\begin{center}
\begin{tabular}{lccccc}
\\
\hline
\hline
Name & RA (J2000.0) & DEC (J2000.0) & $m_{R}$$^{**}$ & $m_{B}$$^{**}$ & 
$d_{S}$ [$\prime$] \\
\hline
B09472 & 15:02:56.05 & $-$41:54:52.5 &  14.9  & 16.2	 & 0.38  \\
B11408 & 15:02:58.25 & $-$41:55:21.0 &  12.5  & 14.4	 & 0.62  \\
B05723 & 15:02:51:85 & $-$41:56:39.6 &  13.7  & 15.3	 & 1.57  \\
B17720 & 15:03:05.35 & $-$41:55:33.2 &  14.8  & 16.0	 & 1.96  \\
B16564 & 15:03:04.07 & $-$41:56:34.7 &  14.3  & 15.6	 & 2.18  \\
B97338 & 15:02:42.59 & $-$41:55:36.5 &  14.7  & 15.5	 & 2.34  \\
B05518 & 15:02:51.59 & $-$41:52:49.3 &  13.8  & 14.8	 & 2.46  \\
B20292 & 15:03:08.22 & $-$41:55:18.7 &  13.5  & 14.9	 & 2.46  \\
B14130 & 15:03:01.31 & $-$41:53:01.3 &  14.5  & 15.6	 & 2.47  \\
B03395 & 15:02:49.27 & $-$41:52:21.9 &  14.8  & 16.2	 & 3.03  \\
B97341 & 15:02:42.59 & $-$41:53:11.0 &  12.9  & 14.9	 & 3.07  \\
B99810 & 15:02:45.34 & $-$41:57:48.3 &  12.3  & 13.4	 & 3.16  \\
B15360 & 15:03:02.70 & $-$41:52:19.3 &  12.6  & 14.6	 & 3.21  \\
B93571 & 15:02:38.42 & $-$41:56:10.7 &  12.4  & 14.1	 & 3.23  \\
B18024 & 15:03:05.69 & $-$41:57:48.0 &  13.7  & 14.1	 & 3.27  \\
B98824$^{*}$& 15:2:44.23 & $-$41:57:50.7 & 13.3 & 14.6   & 3.32  \\
B08277 & 15:02:54.72 & $-$41:51:51.5 &  12.9  & 14.4	 & 3.34  \\
B24215 & 15:03:12.42 & $-$41:54:22.1 &  14.7  & 16.0	 & 3.38  \\
B14707 & 15:03:01.92 & $-$41:51:51.5 &  12.7  & 14.3	 & 3.58  \\
B90474 & 15:02:35.04 & $-$41:54:33.1 &  14.8  & 16.0	 & 3.77  \\
B90102 & 15:02:34.64 & $-$41:55:20.6 &  14.3  & 15.3	 & 3.79  \\
B10074 & 15:02:56.75 & $-$41.51:19.1 &  14.7  & 15.8	 & 3.89  \\ 
B95979 & 15:02:41.10 & $-$41:58:07.8 &  12.9  & 13.6	 & 3.91  \\
B09749 & 15:02:56.39 & $-$41:59:09.0 &  12.7  & 13.7	 & 3.96  \\
B21185$^{*}$& 15:03:09.20 & $-$41:52:12.7 & 13.8 & 15.3  & 3.99  \\      
B26090 & 15:03:14.42 & $-$41:56:55.2 &  12.5  & 13.2     & 4.00  \\
\hline
\end{tabular}
\end{center}

\label{tcoor}

\noindent{*}{Spectroscopic binary}

\noindent{**}{0.03 mag uncertainties}

\end{table}

\clearpage

\begin{table}

\small

{\bf Table S2 \textbar ~Johnson magnitudes $m_{R}$, $m_{B}$, and 
2MASS$^{19}$ magnitudes $m_{J}$, $m_{H}$, $m_{K}$, spectral types and 
luminosity classes$^{46}$, and distances of the sample stars}

\smallskip
\begin{center}
\begin{tabular}{lccccccc}
\\
\hline
\hline
Name & $m_{R}$$^{*}$ & $m_{B}$$^{*}$ & $m_{J}$ & $m_{H}$ & $m_{K}$ &
St.~Type$^{**}$ & $d$ [kpc] \\ 
\hline
B09472 & 14.9 & 16.2 & 14.93$\pm$0.05 & 14.69$\pm$0.07 & 14.55$\pm$0.09 &    G1 V   & 1.23$\pm$0.53 \\
B11408 & 12.5 & 14.4 & 11.05$\pm$0.03 & 10.32$\pm$0.03 & 10.21$\pm$0.02 &  K0-1 III & 1.62$\pm$0.68 \\
B05723 & 13.7 & 15.3 & 13.63$\pm$0.04 & 13.26$\pm$0.05 & 13.19$\pm$0.05 &  G0-1 V   & 0.76$\pm$0.32 \\
B17720 & 14.8 & 16.0 & 13.61$\pm$0.03 & 13.20$\pm$0.03 & 13.12$\pm$0.03 & G9-K0 V   & 0.58$\pm$0.25 \\
B16564 & 14.3 & 15.6 & 12.84$\pm$0.02 & 12.22$\pm$0.03 & 12.13$\pm$0.02 & G9-K0 III & 3.03$\pm$1.27 \\
B97338 & 14.7 & 15.5 & 13.49$\pm$0.02 & 13.17$\pm$0.03 & 13.09$\pm$0.04 &  G4-5 V   & 1.05$\pm$0.45 \\
B05518 & 13.8 & 14.8 & 13.08$\pm$0.03 & 12.79$\pm$0.03 & 12.72$\pm$0.03 &    F6 V   & 0.50$\pm$0.21 \\
B20292 & 13.5 & 14.9 & 12.40$\pm$0.02 & 11.89$\pm$0.02 & 11.83$\pm$0.02 &    G8 IV  & 1.13$\pm$0.48 \\
B14130 & 14.5 & 15.6 & 13.81$\pm$0.03 & 13.49$\pm$0.03 & 13.51$\pm$0.04 &    F7 V   & 0.89$\pm$0.38 \\
B03395 & 14.8 & 16.2 & 13.76$\pm$0.03 & 13.40$\pm$0.03 & 13.37$\pm$0.03 &    G7 V   & 0.95$\pm$0.41 \\
B97341 & 12.9 & 14.9 & 11.94$\pm$0.02 & 11.38$\pm$0.03 & 11.29$\pm$0.02 &    G9 III & 2.48$\pm$1.04 \\
B99810 & 12.3 & 13.4 & 11.59$\pm$0.02 & 10.92$\pm$0.03 & 10.82$\pm$0.03 &    K1 III & 2.42$\pm$1.02 \\
B15360 & 12.6 & 14.6 & 11.43$\pm$0.02 & 10.88$\pm$0.02 & 10.74$\pm$0.02 &    K1 IV  & 1.29$\pm$0.54 \\
B93571 & 12.4 & 14.1 & 10.82$\pm$0.02 & 10.16$\pm$0.03 & 10.01$\pm$0.02 &    K1 III & 2.39$\pm$1.01 \\
B18024 & 13.7 & 14.1 & 12.88$\pm$0.02 & 12.56$\pm$0.03 & 12.50$\pm$0.03 &  F8-9 V   & 0.60$\pm$0.26 \\
B08277 & 12.9 & 14.4 & 12.20$\pm$0.03 & 11.82$\pm$0.02 & 11.71$\pm$0.02 &    G5 V   & 0.45$\pm$0.19 \\
B24215 & 14.7 & 16.0 & 13.91$\pm$0.03 & 13.51$\pm$0.03 & 13.44$\pm$0.04 &    G4 V   & 0.95$\pm$0.41 \\
B14707 & 12.7 & 14.3 & 11.46$\pm$0.02 & 10.98$\pm$0.03 & 10.88$\pm$0.02 &    K0 IV  & 1.37$\pm$0.58 \\
B90474 & 14.8 & 16.0 & 13.43$\pm$0.03 & 12.91$\pm$0.03 & 12.80$\pm$0.03 &    G6 III & 4.78$\pm$2.00 \\
B90102 & 14.3 & 15.3 & 13.30$\pm$0.02 & 12.98$\pm$0.03 & 12.86$\pm$0.03 &    G5 V   & 0.91$\pm$0.39 \\
B10074 & 14.7 & 15.8 & 13.46$\pm$0.03 & 13.09$\pm$0.02 & 12.99$\pm$0.03 &    G6 V   & 0.71$\pm$0.30 \\ 
B95979 & 12.9 & 13.6 & 12.65$\pm$0.02 & 12.52$\pm$0.03 & 12.47$\pm$0.03 &  F1-2 V   & 0.94$\pm$0.40 \\
B09749 & 12.7 & 13.7 & 11.90$\pm$0.02 & 11.60$\pm$0.03 & 11.57$\pm$0.03 &  F7-8 V   & 0.40$\pm$0.17 \\      
B26090 & 12.5 & 13.2 & 12.18$\pm$0.03 & 11.89$\pm$0.03 & 11.86$\pm$0.03 &    F8 V   & 0.37$\pm$0.16 \\
\hline
\end{tabular}
\end{center}

\label{tdist}

\noindent{*}{0.03 mag uncertainties}

\noindent{**}{St.~Type refers to spectral type and luminosity class}

\end{table}

\clearpage

\begin{table}

\small

{\bf Table S3 \textbar ~Photometric distances from different
photometric magnitudes and their averages. The error given for each
individual distance determination of each photometric magnitude 
is the 1-$\sigma$ uncertainty (see text). 
The errors of the average values are the 1-$\sigma$ uncertainties
associated with the dispersion of the mean from the distances of
different magnitudes and the average systematic error, respectively.}

\smallskip
\begin{center}
\begin{tabular}{lcccccc}
\\
\hline
\hline
Name & $d_{m_B}$ & $d_{m_R}$ & $d_{m_J}$ & $d_{m_H}$ & $d_{m_K}$ & 
$d_{\rm av}$ \\   
\hline
 & [kpc] & [kpc] & [kpc] & [kpc] & [kpc] & [kpc] \\   
\hline
B09472 & $1.02\pm0.45$ & $0.95\pm0.41$ & $1.46\pm0.62$ & $1.52\pm0.64$ & $1.46\pm0.62$ & $1.23\pm0.28\pm0.53$ \\
B11408 & $1.75\pm0.75$ & $1.67\pm0.71$ & $1.63\pm0.69$ & $1.54\pm0.64$ & $1.52\pm0.63$ & $1.62\pm0.10\pm0.68$ \\
B05723 & $0.74\pm0.32$ & $0.59\pm0.25$ & $0.86\pm0.36$ & $0.84\pm0.35$ & $0.84\pm0.35$ & $0.76\pm0.11\pm0.32$ \\
B17720 & $0.55\pm0.24$ & $0.60\pm0.26$ & $0.58\pm0.25$ & $0.59\pm0.25$ & $0.58\pm0.25$ & $0.58\pm0.02\pm0.25$ \\
B16564 & $2.78\pm1.19$ & $3.31\pm1.40$ & $3.11\pm1.30$ & $3.00\pm1.24$ & $2.98\pm1.23$ & $3.03\pm0.19\pm1.27$ \\
B97338 & $0.97\pm0.42$ & $1.17\pm0.50$ & $1.04\pm0.44$ & $1.05\pm0.44$ & $1.04\pm0.44$ & $1.05\pm0.07\pm0.45$ \\
B05518 & $0.51\pm0.22$ & $0.50\pm0.21$ & $0.51\pm0.21$ & $0.50\pm0.21$ & $0.50\pm0.21$ & $0.50\pm0.01\pm0.21$ \\
B20292 & $1.09\pm0.47$ & $1.13\pm0.48$ & $1.15\pm0.48$ & $1.13\pm0.47$ & $1.13\pm0.47$ & $1.13\pm0.02\pm0.48$ \\
B14130 & $0.89\pm0.39$ & $0.85\pm0.36$ & $0.90\pm0.38$ & $0.88\pm0.37$ & $0.91\pm0.38$ & $0.89\pm0.02\pm0.38$ \\
B03395 & $0.99\pm0.43$ & $0.94\pm0.40$ & $0.93\pm0.40$ & $0.94\pm0.40$ & $0.96\pm0.41$ & $0.95\pm0.02\pm0.41$ \\
B97341 & $2.59\pm1.10$ & $2.21\pm0.94$ & $2.58\pm1.08$ & $2.55\pm1.05$ & $2.54\pm1.05$ & $2.48\pm0.16\pm1.04$ \\
B99810 & $1.59\pm0.68$ & $2.22\pm0.95$ & $3.08\pm1.30$ & $2.98\pm1.25$ & $2.96\pm1.23$ & $2.42\pm0.67\pm1.02$ \\
B15360 & $1.48\pm0.63$ & $1.23\pm0.52$ & $1.28\pm0.54$ & $1.26\pm0.52$ & $1.23\pm0.51$ & $1.29\pm0.11\pm0.54$ \\
B93571 & $2.34\pm1.00$ & $2.54\pm1.08$ & $2.42\pm1.02$ & $2.38\pm0.99$ & $2.30\pm0.96$ & $2.39\pm0.09\pm1.01$ \\
B18024 & $0.48\pm0.21$ & $0.65\pm0.28$ & $0.65\pm0.27$ & $0.64\pm0.27$ & $0.64\pm0.27$ & $0.60\pm0.07\pm0.26$ \\
B08277 & $0.47\pm0.21$ & $0.42\pm0.18$ & $0.47\pm0.20$ & $0.46\pm0.19$ & $0.45\pm0.19$ & $0.45\pm0.02\pm0.19$ \\
B24215 & $0.96\pm0.42$ & $0.92\pm0.39$ & $0.98\pm0.41$ & $0.96\pm0.40$ & $0.96\pm0.40$ & $0.95\pm0.02\pm0.41$ \\
B14707 & $1.41\pm0.61$ & $1.38\pm0.58$ & $1.36\pm0.57$ & $1.36\pm0.56$ & $1.34\pm0.56$ & $1.37\pm0.03\pm0.58$ \\
B90474 & $4.43\pm1.89$ & $5.22\pm2.21$ & $4.87\pm2.03$ & $4.77\pm1.97$ & $4.70\pm1.94$ & $4.78\pm0.29\pm2.00$ \\
B90102 & $0.84\pm0.37$ & $0.94\pm0.40$ & $0.92\pm0.39$ & $0.94\pm0.40$ & $0.92\pm0.39$ & $0.91\pm0.04\pm0.39$ \\
B10074 & $0.72\pm0.31$ & $0.78\pm0.33$ & $0.69\pm0.29$ & $0.69\pm0.29$ & $0.68\pm0.29$ & $0.71\pm0.04\pm0.30$ \\
B95979 & $0.81\pm0.36$ & $0.84\pm0.36$ & $1.01\pm0.42$ & $1.04\pm0.43$ & $1.03\pm0.43$ & $0.94\pm0.11\pm0.40$ \\
B09749 & $0.40\pm0.18$ & $0.40\pm0.17$ & $0.41\pm0.17$ & $0.40\pm0.17$ & $0.41\pm0.17$ & $0.40\pm0.00\pm0.17$ \\
B26090 & $0.29\pm0.13$ & $0.34\pm0.14$ & $0.43\pm0.18$ & $0.43\pm0.18$ & $0.43\pm0.18$ & $0.37\pm0.07\pm0.16$ \\
\hline
\end{tabular}
\end{center}

\label{tdistb}

\end{table}

\clearpage

\begin{table}

\small

{\bf Table S4a \textbar ~Abundance ratios [X/Fe] of the sample stars 
in the SN~1006 survey for Na and the $\alpha$-elements Mg, Si, Ca 
and Ti. The errors provide the 1-$\sigma$ uncertainties associated
with the dispersion of the measurements from different spectral
features. 
}

\smallskip
\begin{center}
\begin{tabular}{lrrrrr}
\\
\hline
\hline
{Name} & [Na/Fe] & [Mg/Fe] & [Si/Fe] & [Ca/Fe] & [Ti/Fe] \\
\hline 
B09472 & $ 0.30\pm0.08$ & $ 0.20\pm0.09$ & $ 0.14\pm0.18$ & $ 0.06\pm0.21$ & $ 0.17\pm0.18$ \\
B11408 & $ 0.32\pm0.08$ & $ 0.20\pm0.14$ & $ 0.28\pm0.11$ & $ 0.05\pm0.11$ & $ 0.43\pm0.13$ \\
B05723 & $ 0.23\pm0.00$ & $ 0.06\pm0.21$ & $-0.07\pm0.28$ & $-0.03\pm0.09$ & $ 0.08\pm0.03$ \\
B17720 & $ 0.06\pm0.04$ & $ 0.06\pm0.16$ & $ 0.01\pm0.18$ & $-0.11\pm0.07$ & $ 0.11\pm0.17$ \\
B16564 & $ 0.24\pm0.01$ & $ 0.30\pm0.10$ & $ 0.34\pm0.12$ & $ 0.15\pm0.08$ & $ 0.31\pm0.17$ \\
B97338 & $-0.12\pm0.06$ & $ 0.45\pm0.42$ & $-0.03\pm0.11$ & $-0.05\pm0.14$ & $ 0.07\pm0.10$ \\
B05518 & $-0.16\pm0.15$ & $-0.10\pm0.15$ & $-0.11\pm0.08$ & $-0.08\pm0.07$ & $ 0.07\pm0.14$ \\
B20292 & $ 0.08\pm0.05$ & $ 0.30\pm0.26$ & $ 0.30\pm0.09$ & $ 0.19\pm0.06$ & $ 0.36\pm0.09$ \\
B14130 & $ 0.02\pm0.05$ & $ 0.12\pm0.13$ & $ 0.01\pm0.07$ & $-0.05\pm0.10$ & $ 0.02\pm0.13$ \\
B03395 & $ 0.08\pm0.00$ & $ 0.17\pm0.13$ & $ 0.07\pm0.06$ & $-0.03\pm0.04$ & $ 0.13\pm0.12$ \\
B97341 & $ 0.12\pm0.08$ & $ 0.08\pm0.00$ & $ 0.20\pm0.16$ & $ 0.04\pm0.10$ & $ 0.24\pm0.11$ \\
B99810 & $ 0.14\pm0.06$ & $ 0.37\pm0.09$ & $ 0.20\pm0.07$ & $ 0.16\pm0.08$ & $ 0.38\pm0.13$ \\ 
B15360 & $-0.03\pm0.15$ & $ 0.20\pm0.40$ & $ 0.17\pm0.16$ & $-0.04\pm0.09$ & $ 0.14\pm0.15$ \\
B93571 & $ 0.06\pm0.07$ & $ 0.06\pm0.16$ & $ 0.22\pm0.14$ & $ 0.00\pm0.10$ & $ 0.22\pm0.15$ \\
B18024 & $-0.17\pm0.00$ & $ 0.08\pm0.27$ & $ 0.08\pm0.11$ & $ 0.01\pm0.06$ & $ 0.20\pm0.26$ \\
B08277 & $ 0.14\pm0.08$ & $ 0.03\pm0.02$ & $-0.03\pm0.09$ & $-0.03\pm0.07$ & $ 0.04\pm0.06$ \\
B24215 & $ 0.07\pm0.08$ & $ 0.04\pm0.11$ & $ 0.01\pm0.08$ & $-0.00\pm0.09$ & $ 0.11\pm0.15$ \\
B14707 & $ 0.10\pm0.02$ & $ 0.16\pm0.09$ & $ 0.03\pm0.09$ & $-0.03\pm0.03$ & $ 0.09\pm0.14$ \\
B90474 & $ 0.04\pm0.01$ & $ 0.25\pm0.13$ & $ 0.18\pm0.09$ & $ 0.10\pm0.09$ & $ 0.16\pm0.19$ \\
B90102 & $ 0.12\pm0.06$ & $ 0.30\pm0.12$ & $ 0.16\pm0.05$ & $ 0.08\pm0.05$ & $ 0.21\pm0.08$ \\
B10074 & $ 0.04\pm0.03$ & $-0.03\pm0.01$ & $ 0.01\pm0.09$ & $ 0.02\pm0.08$ & $ 0.10\pm0.14$ \\
B95979 & $ 0.11\pm0.09$ & $-0.24\pm0.00$ & $ 0.09\pm0.06$ & $-0.05\pm0.07$ & $ 0.15\pm0.18$ \\
B09749 & $ 0.07\pm0.00$ & $ 0.04\pm0.13$ & $-0.05\pm0.11$ & $-0.06\pm0.14$ & $ 0.22\pm0.23$ \\
B26090 & $-0.00\pm0.02$ & $-0.16\pm0.06$ & $-0.03\pm0.07$ & $-0.00\pm0.07$ & $ 0.03\pm0.06$ \\
\hline
\end{tabular}
\end{center}

\label{taba}

\end{table}


\clearpage

\begin{table}

\small

{\bf Table S4b \textbar ~Abundance ratios [X/Fe] (cont.)
of the sample stars in the SN~1006 survey
for the iron-peak elements Cr, Mn, Co and Ni.}

\smallskip
\begin{center}
\begin{tabular}{lrrrr}
\\
\hline
\hline
{Name} & [Cr/Fe] & [Mn/Fe] & [Co/Fe] & [Ni/Fe] \\
\hline
B09472 & $ 0.14\pm0.21$ & $-0.10\pm0.10$ & $ 0.03\pm0.21$ & $ 0.07\pm0.12$ \\
B11408 & $-0.00\pm0.15$ & $-0.03\pm0.21$ & $ 0.28\pm0.10$ & $ 0.08\pm0.11$ \\
B05723 & $-0.09\pm0.11$ & $ 0.22\pm0.22$ & $ 0.17\pm0.12$ & $ 0.03\pm0.19$ \\
B17720 & $-0.11\pm0.09$ & $ 0.22\pm0.31$ & $ 0.17\pm0.12$ & $ 0.07\pm0.10$ \\
B16564 & $-0.03\pm0.16$ & $ 0.01\pm0.10$ & $ 0.38\pm0.18$ & $ 0.09\pm0.10$ \\
B97338 & $-0.08\pm0.12$ & $ 0.00\pm0.09$ & $-0.01\pm0.13$ & $-0.03\pm0.12$ \\
B05518 & $-0.11\pm0.08$ & $-0.04\pm0.16$ & $-0.19\pm0.13$ & $ 0.00\pm0.12$ \\
B20292 & $ 0.00\pm0.15$ & $-0.15\pm0.15$ & $ 0.18\pm0.06$ & $ 0.03\pm0.10$ \\
B14130 & $-0.03\pm0.08$ & $ 0.12\pm0.16$ & $-0.01\pm0.09$ & $ 0.06\pm0.07$ \\
B03395 & $-0.07\pm0.05$ & $ 0.14\pm0.29$ & $ 0.17\pm0.06$ & $ 0.07\pm0.07$ \\
B97341 & $-0.04\pm0.23$ & $ 0.07\pm0.09$ & $ 0.27\pm0.12$ & $ 0.03\pm0.10$ \\
B99810 & $-0.01\pm0.18$ & $-0.15\pm0.08$ & $ 0.15\pm0.14$ & $ 0.04\pm0.12$ \\
B15360 & $-0.06\pm0.10$ & $ 0.14\pm0.21$ & $ 0.32\pm0.28$ & $ 0.09\pm0.12$ \\
B93571 & $ 0.04\pm0.08$ & $-0.18\pm0.20$ & $ 0.23\pm0.17$ & $ 0.04\pm0.15$ \\
B18024 & $-0.01\pm0.27$ & $-0.12\pm0.12$ & $ 0.05\pm0.29$ & $-0.05\pm0.13$ \\
B08277 & $-0.02\pm0.07$ & $ 0.14\pm0.13$ & $ 0.10\pm0.13$ & $ 0.03\pm0.06$ \\
B24215 & $-0.06\pm0.12$ & $ 0.02\pm0.07$ & $ 0.11\pm0.11$ & $ 0.06\pm0.11$ \\
B14707 & $-0.09\pm0.12$ & $ 0.31\pm0.11$ & $ 0.14\pm0.12$ & $ 0.01\pm0.13$ \\
B90474 & $-0.11\pm0.14$ & $-0.02\pm0.16$ & $ 0.23\pm0.19$ & $-0.00\pm0.09$ \\
B90102 & $-0.04\pm0.14$ & $-0.14\pm0.05$ & $ 0.14\pm0.12$ & $ 0.04\pm0.09$ \\
B10074 & $-0.04\pm0.09$ & $ 0.06\pm0.08$ & $ 0.06\pm0.11$ & $ 0.03\pm0.11$ \\
B95979 & $-0.06\pm0.19$ & $-0.09\pm0.25$ & $-0.11\pm0.27$ & $-0.05\pm0.23$ \\
B09749 & $ 0.05\pm0.16$ & $-0.06\pm0.20$ & $ 0.05\pm0.21$ & $ 0.06\pm0.09$ \\
B26090 & $-0.05\pm0.10$ & $ 0.03\pm0.05$ & $-0.05\pm0.10$ & $-0.02\pm0.07$ \\
\hline
\end{tabular}
\end{center}

\label{tabb}

\end{table}

\end{document}